\documentclass[twocolumn]{revtex4-1}
\pdfoutput=1
\usepackage{graphicx}
\usepackage{natbib}
\begin{document}

\title{Control and Manipulation of Cold Atoms in Optical Tweezers}

\author{Cecilia Muldoon}
\author{Lukas Brandt}
\author{Jian Dong} 
\author{Dustin Stuart}
\author{Edouard Brainis}
\altaffiliation{now at: Service OPERA (CP 194/5), Universit\'{e} libre de Bruxelles, Avenue F.\,D. Roosevelt 50, 1050 Brussels, Belgium}

\author{Matthew Himsworth}
\altaffiliation{now at: School of Physics and Astronomy, University of Southampton, SO17 1BJ, UK}

\author{Axel Kuhn}
\email{axel.kuhn@physics.ox.ac.uk}
\affiliation{University of Oxford, Clarendon Laboratory, Parks Road, Oxford, OX1 3PU, UK}

\date{\today}

\begin{abstract}
Neutral atoms trapped by laser light are amongst the most promising candidates for storing and processing  information in a quantum computer or simulator. The application certainly calls for a scalable and flexible scheme for addressing and manipulating the atoms. We have now made this a reality by implementing a fast and versatile method to dynamically control the position of neutral atoms trapped in optical tweezers. The tweezers result from a spatial light modulator (SLM) controlling and shaping a large number of optical dipole-force traps. Trapped atoms adapt to any change in the potential landscape, such that one can re-arrange and randomly access individual sites within atom-trap arrays. 
\end{abstract}

\maketitle

The ability to address  and manipulate individual information carriers in a deterministic and scalable manner is a central theme in quantum information science \cite{DiVincenzo00}. Examples of matter-based quantum systems that could serve as information carriers include trapped ions \cite{Benhelm08, Myerson08, Singer10, NIST02}, magnetically trapped atoms \cite{AtomChips,Folman02} and dipole-trapped neutral atoms \cite{Alt03, Schlosser01, Volz06}. The latter are very appealing as the absence of magnetic trapping fields implies no Zeeman shifts and no spin precession, such that the atomic states can be manipulated precisely and preserve coherence. Also, the insensitivity to the electrostatic environment and the largely absent interaction between atoms allow for large-scale atomic arrays close to dielectric surfaces, such as cavity mirrors or hollow-core fibres. In single optical dipole-force traps, the manipulation of individual atoms has been explored extensively \cite{Beugnon07, Karski09, Grunzweig10}, and the scalability of dipole trapping has been shown in optical lattices \cite{Sherson10, Bakr09, Weitenberg11}, as well as in dipole-trap arrays created by artificial holograms using spatial phase modulators \cite{Bergamini04, Boyer06}, acousto-optic devices \cite{Fatemi07,Henderson09}, and micro-lens arrays \cite{Kruse10, Lengwenus10}. 

Here, we show how to combine the scalability of these lattices and arrays with the flexibility of controlling individual trapping sites. We confine neutral atoms in arbitrarily shaped and easily reconfigurable trapping potentials which we generate using the  direct image of a  Digital Mirror Device (DMD) \cite{MacAulay98, Hanley99}. In contrast to other schemes, our approach allows for the  fast and independent manipulation and control of a large number of trapping sites. In particular, the DMD attains frame rates which are much faster than our typical trapping frequencies. This is a considerable advance compared to holographic phase modulation techniques that are inherently slow and thus unable to act on the timescale required for successful atom transport. We investigate how trapped atoms adapt to changes in the potential landscape, and use this to move them to arbitrary positions. The latter is accomplished with a ballistic `release and recapture' scheme in which atoms are transported with minimal heating.

\section{Trap design and optical arrangement}
\begin{figure*}[t]
\includegraphics[width=0.9\textwidth]{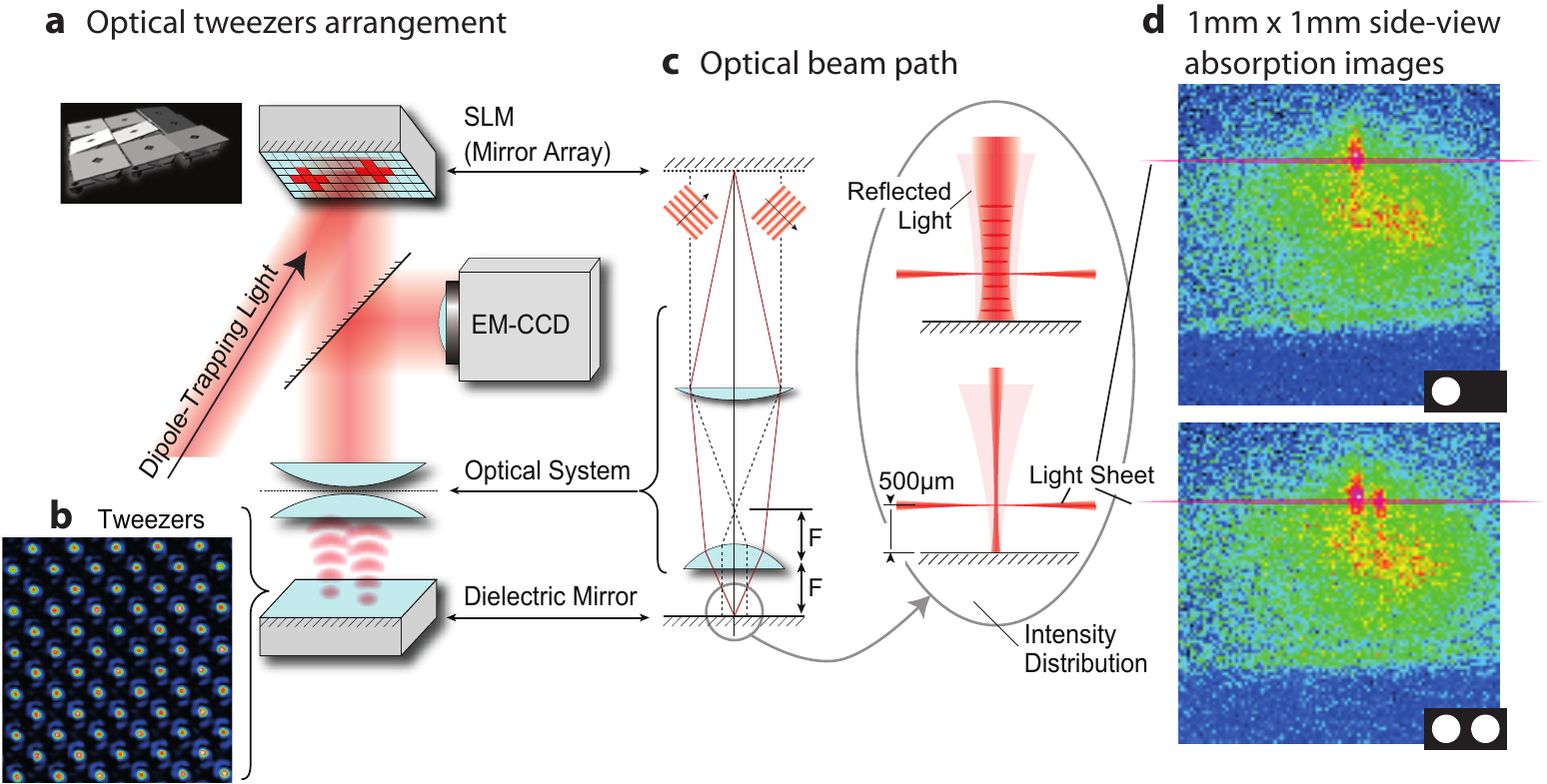}
\caption{\textbf{Tweezers arrangement.} 
\textbf{a}, Dipole trapping light illuminates a SLM and is focused through a two lens microscope to form arbitrary potential landscapes which can be used to trap and guide atoms. A thin sheet of resonant light is used to illuminate only the atoms trapped in the image plane, and a highly sensitive EMCCD camera is used to observe the atomic fluorescence via a dichroic mirror. \textbf{b}, Tweezers array of $5\,\mu$m spacing and $1\pm0.2\,\mu$m trap waist. Depicted is the false-colour intensity profile of the trapping light ranging from zero (dark blue) to $30\,\mu$W/$\mu$m$^2$ (bright red). \textbf{c} The two lens microscope used to image the surface of the SLM consists of an achromatic doublet and a high numerical aperture aspheric lens.The close-up view illustrates that interference between the incoming and reflected light leads to a standing wave, and how this effect is negligible for smaller traps. \textbf{d}, Side-view false-colour absorption images of atoms in either one or two dipole traps of $\sim30\,\mu$m radius each. The  snapshots are taken immediately after loading, with a diffuse cloud of untrapped atoms from the MOST still surrounding the traps  (blue: no absorption; red: maximum absorption and highest atom density).}
\label{setup}
\end{figure*}

The work presented here is the first experimental implementation of the atom trapping and manipulation scheme that we recently proposed theoretically \cite{Brandt11}. Reconfigurable optical dipole-force traps, or optical tweezers, are produced by imaging the surface of a spatial light modulator onto a cloud of cold $^{87}$Rb atoms (see Fig.\,\ref{setup}). The atoms are confined in the potential caused by the resulting intensity distribution in the image plane. This potential takes the form
\begin{equation}
U_{dip}\approx\frac{\hbar\Omega^2}{4\delta}
\label{eq1}
\end{equation}
where $\Omega$ is the Rabi frequency, and $\delta$ is the detuning of the trapping laser with respect to the closest allowed atomic transition. It results from the AC Stark shift induced by the applied field, and its gradient gives rise to a conservative force known as the dipole force \cite{Foot05}. If $\delta<0$ (i.e. the trapping beam is red-detuned), this force acts in the direction of increasing intensity, and $U_{dip}$ is attractive: this is what is referred to as a dipole-force trap. In the present work, we confine $^{87}$Rb atoms in potential wells that are at most  $k_B\times 100\,\mu$K deep (see Methods). To load atoms into these conservative traps, we first capture and cool them from the background gas. This is accomplished with a Magneto Optical Surface Trap (MOST) \cite{Wildermuth04} providing a $\sim280\,\mu$K cold cloud of $\sim120\,\mu$m radius containing approximately $10^{5}$ atoms at 500$\,\mu$m distance from a mirror surface.

The SLM is an array of 1024$\times$768 independently addressable micro mirrors (see Methods). Its surface is imaged onto the atoms using the two-lens microscope shown in Fig.\,\ref{setup}a. For large traps, the divergence of the trapping beam is small and interference between the incoming and reflected light leads to a standing wave. This is advantageous as in addition to axial confinement, the atoms will also experience some degree of longitudinal confinement. Atoms that are cold enough are therefore confined to anti-nodal planes stacked along the optical axis. This effect only partially applies to smaller traps (of radii  $\lesssim 20\,\mu$m) because the reflected beam diverges too quickly. The microscope has a (de)magnification of 57:1, which means that individual micro mirrors are not resolved. A diffraction limited spot of 1$\,\mu$m corresponds to a block of 4$\times$4 mirrors, which implies that we can mimic up to 16 levels of intensity by dithering the mirror pattern on the otherwise digital SLM. Also, taking both the finite dimension of the SLM and the resolution limit into account, a two-dimensional array of up to $50\times 40$ well-separated individual trapping sites can be formed. The trapping-light intensity profile depicted in Fig.\,\ref{setup}b is an excerpt of such an array, with a distance between traps equal to five times the resolution limit.  

Atoms are detected by fluorescence imaging through the same microscope used to produce the tweezers. The atoms in the image plane of the tweezers are illuminated by a 9.7$\,\mu$m thin sheet of resonant light parallel to the mirror surface, while atoms above or below remain in the dark. The light sheet is 5.1\,mm wide and is retro-reflected to avoid pushing the atoms out of the traps. It is blue-detuned by 2\,MHz from the 5$^{2}\textit{S}_{1/2}\leftrightarrow5^{2}\textit{P}_{3/2}$ cycling transition in order to account for the tweezers-induced dynamic Stark shift. The total power in the light sheet is ~7$\,\mu$W, which gives rise to a scattering rate of 16 photons$/\mu$s. The fluorescence light from the atoms is separated from the reflected trapping light using a dichroic filter and directed to a sensitive camera to record fluorescence images. Additional band pass filters in front of the camera eliminate the unwanted trapping light. The camera's exposure window is synchronised with timing of the light sheet. 

\begin{figure*}[t]
\includegraphics[width=0.9\textwidth]{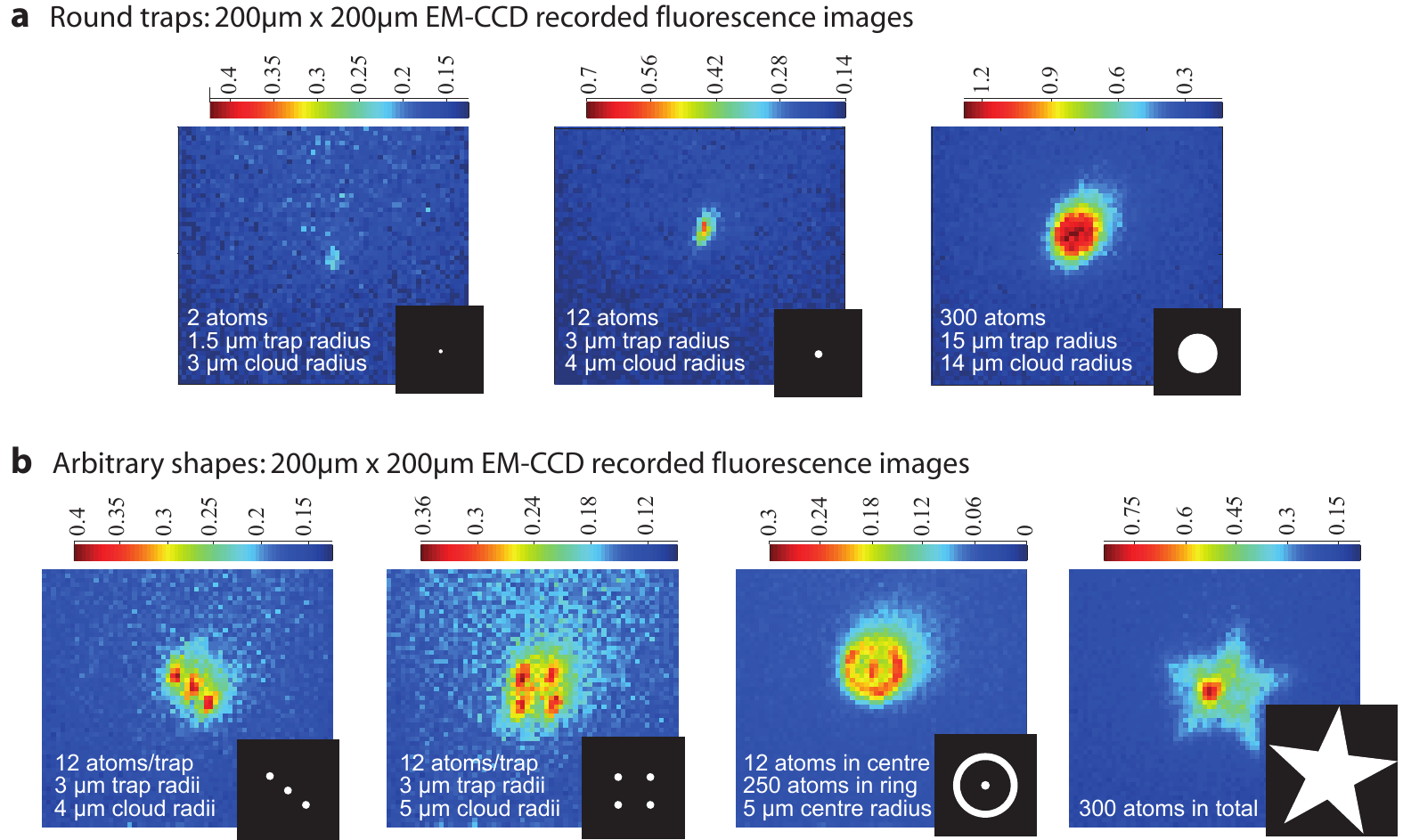}
\caption{\textbf{Fluorescence images of trapped atoms.} \textbf{a}, Single static round traps of different sizes. \textbf{b}, Atoms in various arbitrary trapping geometries: a line, grid, bullseye and star. Each image is an average over 20 consecutive trap loading cycles. The colour scale has units of atoms per $\mu$m$^2$. The actual patterns on the SLM used to produce the traps are shown in the respective insets using the same length scale. The halo seen in the first three fluorescence images can be attributed to diffusion of the atoms in the light sheet.}
\label{shapes}
\end{figure*}

\section{Atoms trapped in arbitrary potential landscapes} 
The top view through the microscope is well-suited for arbitrarily manipulating and observing atoms in a two-dimensional plane. However, atoms are in general trapped in a vertical column perpendicular to the image plane. We monitor this from the side by absorption imaging. Fig.\,\ref{setup}d shows a particular situation with atoms loaded into either one or two trapping beams of $30\,\mu$m radius. Obviously, atoms get transferred into the standing-wave dipole trapping beams at various heights above the mirror, so that observing all atoms by fluorescence imaging through the microscope would normally give rise to blurry pictures. As we are primarily interested only in those atoms trapped in or near to the image plane, we resolved that issue using the above-mentioned light sheet to avoid excitation of atoms being out-of-focus. Independent optical characterisation of our imaging system (see Fig.\,\ref{setup}b) shows that the imaging system has a resolution of 1$\pm0.2\,\mu$m, which also limits the edge sharpness of any trapping potential.

The fluorescence images taken through the microscope reveal that the shape of the atom cloud adapts to the imposed trapping geometry.  Fig.\,\ref{shapes}a shows atoms confined to a series of circular flat-bottom traps. The traps are created by displaying disks of different radii on the homogeneously illuminated SLM. Normally, one would expect the atoms to fill the trap, such that the radius of the atom cloud equals the trap radius. This is indeed found for traps of large size, which demonstrates that the atoms adapt to the imposed trapping geometries. Deviations are only found for very small traps, where some atoms trapped away from the image plane appear out of focus. 

The approximate atom number in each trap was deduced from fluorescence images averaged over 20 loading cycles. During each cycle, the light sheet is flashed on for 90$\,\mu$s, within which a single atom emits $\sim1400$ photons; of these, 21 photons are detected (see Methods for the collection efficiency). We found that the large trap ($\O\,30\,\mu$m) contained 
300$\pm40$ atoms, the 
medium trap ($\O\,6\,\mu$m) 12$\pm2$ atoms, and the
smallest observed trap ($\O\,3\,\mu$m) 2.0$\pm0.3$ atoms. This is very close to the single-atom regime, which one might reach by application of the collisional blockade technique \cite{Kuppens00, Schlosser01} (a widely used method for loading dipole traps with at most one atom).   

The lifetime of atoms in all traps is approximately 50\,ms, and does not change significantly with the size of the trap. For all traps of radius $\ge 3\,\mu$m, the atom number was sufficiently large for time-of-flight measurements, which yield a consistent temperature of $90\pm10\,\mu$K. This is on the same scale as the calculated trap depth. 

We furthermore verified the versatility of the tweezers by trapping atoms in arbitrary shaped potentials. The trap arrays shown in Fig.\,\ref{shapes}b underline the scalability of the scheme: such arrays could be used to implement a large register of atomic qubits. In the first two images, the radius of each individual trap is 3$\,\mu$m, and the average atom number in each is 12$\pm2$. The third image shows atoms trapped in a bullseye-shaped trap. The $\O\,6\,\mu$m centre of the bullseye contains 12$\pm2$ atoms, whilst the ring contains  250$\pm33$ atoms. This ring structure shows that it should be possible to transport atoms not just along straight paths, but also along curved trajectories. 

The last image in Fig.\,\ref{shapes}b shows 300$\pm40$ atoms in a star-shaped trap which is affected by an inhomogeneity in the trapping light. The intensity variation of the laser over the whole area of the SLM is about 30\%. This has no significant impact on smaller traps. However, in larger flat bottomed traps like the star, it leads to a congregation of atoms in the regions of highest intensity. Notwithstanding, the large variety of different shapes highlights the flexibility of trap generation with the SLM. Trapped atoms can be arranged into arbitrary patterns using the SLM to impose any desired shape of the trapping potential. 

For the larger traps depicted in Fig.\,\ref{shapes}, the atom number stated is always restricted to those atoms exposed to the light sheet due to diffusion of the hottest atoms out of the trap. For trap radii $\leq3\,\mu$m, confinement along the vertical is dominated by the waist of the trapping beam, so the atom number measured equals the real atom number in the trap. The density of atoms and the small size of the traps leads to a maximum reduction of the fluorescence of $1\%$ due to reabsorption, and hence does not lead to an underestimation of the observed atom number.

\begin{figure}
\includegraphics[width=\columnwidth]{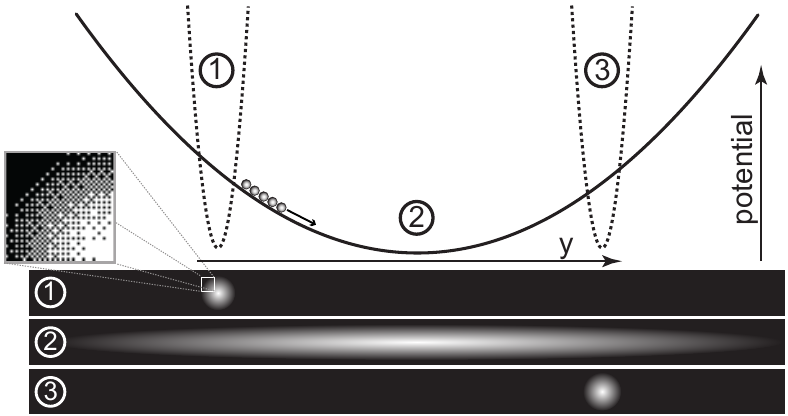}
\caption{\textbf{Illustration of the concept of the atom transport} via ballistic `release and recapture' (not to scale). The atoms are first confined in trap (1), then move along the harmonic transport trap (2) and refocus at the destination when the end trap (3) is  switched on. The close-up view shows the dithering of the mirror pattern we apply to obtain a smooth trapping potential.}
\label{scheme}
\end{figure}

\begin{figure*}
\includegraphics[width=\textwidth]{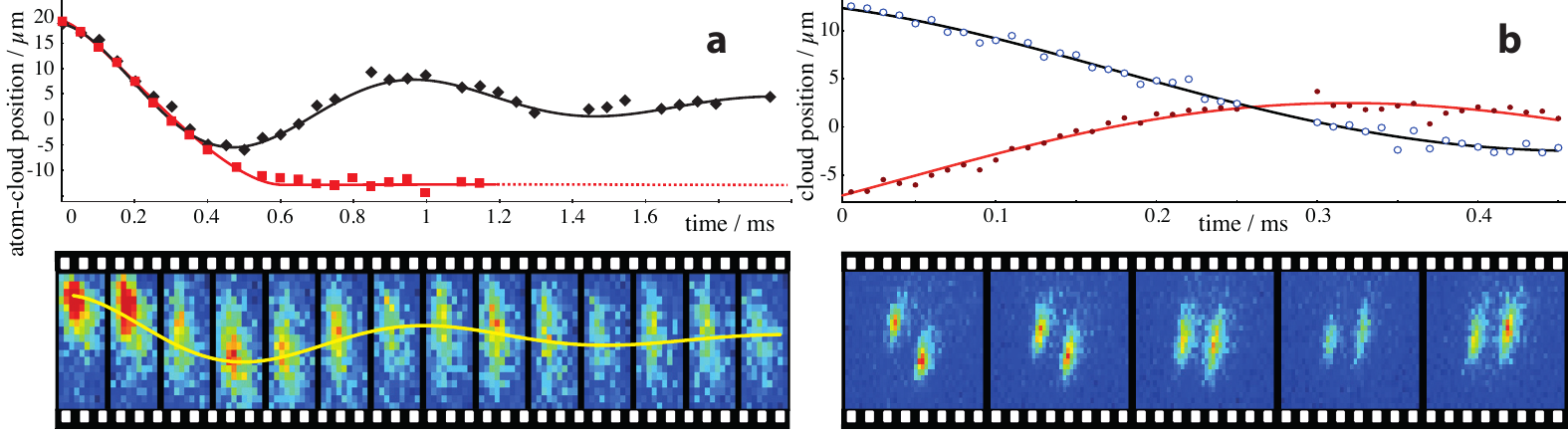}
\caption{\textbf{Ballistic atom transport}. The harmonic transport traps, $U \propto -(1-(x/w_x)^2-(y/w_y)^2)$, are $w_{x} = 6\,\mu$m wide and $w_{y} = 26\,\mu$m wide, whilst the round harmonic start and end traps all have a waist of $w_{r} = 6\,\mu$m. \textbf{a}, Oscillation (black trace), and recapture (red trace) of atoms. The filmstrip below shows a sequence of fluorescence images taken at various instants of the transport.  \textbf{b}, Atoms oscillating past each other in two separate harmonic wells. }
\label{transport}
\end{figure*}

\section{Deterministic re-arrangement and controlled transport of atoms}
So far, we have shown that atoms can be held in a variety of arbitrarily shaped traps. This includes regular arrays of atoms, which one could use as quantum registers in a scalable quantum processor. However, to fill such an array with preferentially one atom per site, or to arbitrarily access a random cell within  a register, it is necessary to move atoms independently from one trapping site into another. Here, we show how to accomplish this task and deterministically transport atoms between two well defined positions.  The most intuitive way for doing so would be to move the trap of interest by gradually changing the image on the SLM such that the centre of the trap gets displaced. However, this approach is hampered by the discrete switching of the SLM's mirrors, which results in the trap hopping along its trajectory rather than in a smooth motion. In turn, the atoms are lost due to excess heating. Our solution therefore is to let the atoms roll along a harmonic potential before recapturing them in the outer turning point. This ballistic `release and recapture' transport effectively entails opening an elongated harmonic trapping channel between a start and end position, and timing the recapture to coincide with half the oscillation period of the atoms in this potential well. Fig.\,\ref{scheme} illustrates that procedure and also shows the dithered mirror pattern used to obtain a  smooth trapping potential in the image plane. 

In general, one has to take into account that the atoms move in three dimensions. The SLM controls the potential in the trapping plane, while the confinement perpendicular to it is controlled by a combination of the near-field propagation and standing-wave effects resulting from the interference between the trapping light and its own reflection. 
We calculated the near-field intensity distribution away from the image plane using the Fresnel-Kirchhoff diffraction integral.
The divergence is moderate enough to yield a standing-wave modulation visibility of 75\% at the height of the image. This is very strong, and most atoms are confined to the resulting anti-nodal planes. Therefore, we effectively have 20 identical two-dimensional trapping planes within the width of the light sheet. 
%Therefore the observed atoms are  subject to identical potentials within the limited width of the light sheet we use for fluorescence detection.

Fig.\,\ref{transport}a shows atoms oscillating along the transport channel (black trace) and illustrates that we are capable of recapturing them at the turning point after half a period of oscillation (red trace). The displayed atom-cloud position is the centre-of-mass of the recorded  fluorescence images, which were taken at constant rate. Of the 22 atoms initially in the  trap, 12 atoms are recaptured. Most surprisingly, this cannot be attributed to heating-induced losses, as time-of-flight measurements reveal a temperature rise of only 10$\,\mu$K. Instead, the reduction in fluorescence is to some extent explained by atoms not exactly regrouping at the destination. The primary reason for that limitation is the small anharmonicity of the transport trap, resulting in a 1.3\,ms damping time of the harmonic motion (see Fig.\,\ref{transport}a, black trace).  The secondary loss mechanism is a diffusion of the most energetic atoms along the standing wave, i.e. away from the image plane. Both losses could be easily reduced using a more powerful trapping laser. Hence disregarding the spreading, the successful ballistic  transport in harmonic traps illustrates that the SLM is an ideal tool for positioning atoms in a deterministic way.

The manipulation of individual information carriers in a large-scale quantum network relies on the capability of re-arranging atoms sitting at different nodes of an array independently from each other. To demonstrate that this can be done, we use two transport channels to move atoms in opposite directions from two different starting positions. We apply the same release and recapture method as discussed above. Fig.\,\ref{transport}b shows atoms oscillating past each other in these harmonic potentials. The atoms move to destinations 8$\,\mu$m and 9$\,\mu$m away from their respective starting  points, with the two destinations chosen independently. This individual transport of randomly selected trapping sites shows that SLM-controlled atom traps are capable of regrouping arrays of trapped atoms to nearly any arbitrary pattern.

\section{Conclusion and Outlook}
SLM-based optical-tweezers constitute a flexible scheme for the manipulation and transport of dipole-trapped neutral atoms.  Individual trapping sites within a large array can be controlled independently, which is essential for a scalable system. The refresh rate of the SLM is fast enough for a dynamic control of trapped atoms in response to the observed fluorescence, such that a feed-back scheme could be realised to control the atom number in individual sites. 
With the ballistic transport applied to individual atoms, one might, e.g.,  move two atoms into a fibre-tip micro cavity and realise pair-wise entanglement \cite{Marr03}, or use controlled collisions between two atoms to implement a two-qubit gate \cite{Jaksch99,Weitenberg11_2,Saffman10}. 
Hence scalable atomic arrays for quantum computing or simulation are now in reach of current technology.

\section{Methods}
\textbf{The spatial light modulator} 
(SLM -- Texas Instruments DMD Discovery 1100) consists of 1024 by 768 independently addressable micro-mirrors, 13.7$\,\mu$m by 13.7$\,\mu$m each, that can be digitally switched between two tilt angles corresponding to an on and off position, the on position directing the impinging light onto the desired optical axis, and the off position throwing it onto a beam dump. These mirrors can be controlled individually, allowing us to generate traps of virtually any shape. The entire optical surface covered by these mirrors has a size of 14\,mm by 10.5\,mm. The SLM can display either static patterns or movies made from frames of these patterns. The movie frame rate ranges between 4 and 20\,KHz, making it fast enough for a dynamic control of the atoms trapped in the tweezers. 

\textbf{Trap depth:} 
The SLM is illuminated using a high power diode laser (Toptica DLX-110). This laser has a power output of 800\,mW and is red-detuned by 5\,nm ($\lambda$=785\,nm) from the 5$^{2}\textit{S}_{1/2}\leftrightarrow5^{2}\textit{P}_{3/2}$ transition in $^{87}$Rb. We have chosen this relatively small detuning to achieve a reasonable trap depth at light intensities below the damage threshold of the SLM. The lens distances and convergence of the illuminating beam have been chosen such that isoplanatism is ensured \cite{Brainis09}, i.e. the wavefront in the image plane is identical to the wavefront of the light in the plane of the SLM. The beam illuminating the SLM has a Gaussian profile that fills the area of interest (typically 1/3 of the SLM surface). In the image plane, the intensity of the trapping laser reaches $30\,$W/mm$^2$, which corresponds to a maximum potential depth of $k_B\times 100\,\mu$K.

\textbf{Optical system:}
A two-lens microscope images the surface of the SLM onto a plane 500$\,\mu$m above the surface of a mirror. We use a high numerical aperture aspheric lens so that our optical tweezers have the smallest possible foci, and hence the best possible axial confinement. The microscope consists of this aspheric lens, which has a focal length of 20\,mm and a numerical aperture of 0.52, and an achromatic doublet as a collimator, with a focal length of 750\,mm. Only the aspheric lens is mounted inside the vacuum chamber. An important feature of this system is that the same two lenses used to image the SLM are also used to observe the atoms trapped in the tweezers by laser-induced fluorescence, as illustrated in Fig.\,1. A dichroic mirror is used to separate the trapping light from the collected atomic fluorescence at 780\,nm. An electron-multiplying CCD camera (EMCCD) is used to observe the laser-induced fluorescence (Andor iXon 885).

\textbf{Collection efficiency:}
The  photon-detection efficiency of the system was found to be $\eta_{tot}=(1.5\pm0.2)\%$, where $\eta_{tot}$ is the product of the collection efficiency of the lens system $\eta_{lens}$=7.6$\%$, the losses along the optical path $\eta_{loss}$=51$\%$, and the quantum efficiency of the camera $\eta_{QE}$=41$\%$.

%\bibliographystyle{unsrt}
%\bibliography{../../../literature/axelsbib.bib}

\section*{Acknowledgements}
This work was supported by the Engineering and Physical Sciences Research Council (EP/E023568/1), the Deutsche Forschungsgemeinschaft (DFG, RU 635), the Philippe Wiener and Maurice Anspach Foundation, and the EU through the RTN EMALI (MRTN-CT-2006-035369). E.B. acknowledges support by the Belgian Fonds de la Recherche Scientifique - FNRS.

\end{document}